\documentstyle[multicol,aps,psfig]{revtex}
\begin{document}

\draft

\preprint{\vbox{\hbox{DRAFT of U. of Iowa preprint 2000-2505}}}

\title{A Simple Method to  Make Asymptotic Series 
of Feynman Diagrams Converge}
\author{Y. Meurice \\ 
{\it Department of Physics and Astronomy, The University of Iowa, 
Iowa City, Iowa 52242, USA}}

\maketitle
\begin{abstract}
We show that for two non-trivial $\lambda \phi ^4$ problems (the 
anharmonic oscillator and the Landau-Ginzburg hierarchical model), 
improved perturbative series can be obtained 
by cutting off the large field contributions.
The modified series converge to values exponentially close to the exact ones.
For $\lambda$ larger than some critical value,
the method outperforms Pad\'e's approximants and Borel summations. 
The method can also be used for series which are not Borel summable 
such as the double-well potential series. We show that
semi-classical methods can be used to calculate the modified
Feynman rules, estimate the error and optimize the field cutoff.

\end{abstract}

\pacs{PACS: 11.10.-z, 11.15.Bt, 12.38.Cy, 31.15.Md}
\begin{multicols}{2}\global\columnwidth20.5pc
\multicolsep=8pt plus 4pt minus 3pt
 
Perturbative series associated with Feynman diagrams are commonly used 
in particle physics, solid state physics, optics
and chemistry \cite{feynman85}. 
One remarkable success of this method is 
the prediction of the values of the anomalous magnetic 
moments of the electron and the muon with an incredible accuracy.
Perturbative methods 
are also used to 
perform precision tests of the standard model of electro-weak and strong 
interactions \cite{sirlin99}.
Despite these successes, it has been known for a long time 
\cite{dyson52,leguillou90} 
that series calculated from Feynman diagrams are not convergent but 
asymptotic. In other words, the range of validity shrinks with the order.
One can improve this situation
by using Pad\'e approximants \cite{baker96}, either on the original series or
a Borel sum \cite{baker76b} of the series, if meaningful. 
However, even in the cases where
the convergence of these alternate procedures can be proven, the convergence
is very slow when the coupling is too large. In addition, for short series,
it is difficult to estimate the error and to choose 
the best approximants.

In this Letter, we construct improved perturbative series which converge
to values which are exponentially close to the exact ones. The error 
can be estimated analytically. 
The method can be applied on the lattice and in the 
continuum and works well 
when the methods mentioned above are inefficient or not applicable.
The method is a perturbative version of recent numerical
calculations performed for various 
$\lambda \phi ^4 $ models,  namely the anharmonic oscillator\cite{bacus} and 
the Landau-Ginzburg model in the hierarchical
approximation \cite{gam3rapid}. In these calculations, we were led to 
introduce large $\phi$  cutoffs and realized that as $\lambda$ increases,
the field cutoff can be decreased without affecting the accuracy of the 
result. The calculations presented here are perturbative series 
calculated with a large field cutoff.
We only consider $\lambda \phi ^4$
problems, however the procedure should extend to 
any kind of model where large field configurations are suppressed
at positive coupling. This is 
in agreement with the general argument \cite{pernice98} 
linking the large field configurations to
the impossibility of applying Lebesgue dominated convergence to 
the path integral expression of the perturbative series.

\begin{figure}
\centerline{\psfig{figure=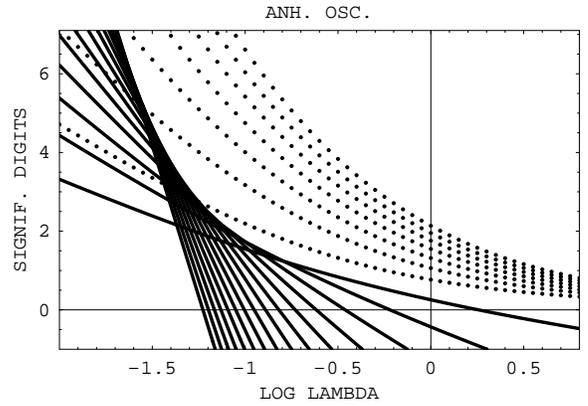,width=3.in}}
\vskip10pt
\caption{
Number of correct significant digits
obtained with regular perturbation theory (solid lines) 
at order 1, 2, 3, ..., 15 and with
Pad\'e approximants [2/2], [3/3], .... 
[7/7] (dots)
for the anharmonic oscillator, vs. log$_{10}\lambda$. 
The various orders can be identified from the explanations 
in the text.
In all the graphs, the logarithms are in base 10.}
\label{fig:one}
\end{figure}

In order to give an idea about the efficiency of existing methods,
we consider the well-known 
example of 
the 
ground state energy of the anharmonic oscillator 
\cite{bender69}. 
The solid lines of 
Fig. \ref{fig:one} represent the number of significant digits obtained
with perturbation theory for various $\lambda$.
As the order increases, the approximate lines
rotate clockwise while moving left, forming an approximate envelope.
For a fixed coupling, there is an order of perturbation for which the error
is minimized. 
On the other hand, the number of digits obtained with 
Pad\'e approximants increases with the order. This
is a consequence of Carleman's theorem which can be used to show
\cite{loeffel69} 
that diagonal sequences
of Pad\'e approximants converge to the ground state energy 
in an appropriately restricted domain of 
the complex plane. However, the convergence rate becomes slower as the 
coupling increases. This can be explained \cite{loeffel69} from the 
fact that, when the coupling increases, 
a $[L/L]$ approximant tends to a constant while the energy increases
like $\lambda^{1/3}$. 
If Pad\'e approximants are used for the Borel transform instead 
of the series, one
obtains results qualitatively similar which are discussed later.
The method that we propose provides a systematic and 
controllable improvement to regular perturbation and a better convergence 
than the Pad\'e based methods in the right hand part
of Fig. \ref{fig:one}. It can also be used in cases where the Borel sum 
has singularities on the positive real axis.
In particle physics, the experimental error bars 
of precision measurements
are shrinking and higher
order terms of perturbative series are being calculated. 
Firmly established 
discrepancy or agreement between theory and experiment provides 
valuable information regarding the laws of nature at shorter distances.
A common practice \cite{ellis96} to estimate at which 
order, for a given coupling, we reach the envelope
illustrated in Fig. \ref{fig:one},  
is to determine when the 
ratio of successive contributions reaches one. 
This method works quite accurately for the three examples considered
below.
It seems reasonable to interpret ratios of 
successive contributions close to one as a signal 
that one needs to go beyond regular 
perturbative theory. We are getting close to this situation.
For instance,
the electro-weak corrections to $g_{\mu}-2$, give 
a contribution \cite{marciano96}
of $151(4)\times10^{-11}$ and in this calculation, the two-loop effects
reduce the one-loop prediction by 35 percent.
The total electroweak contribution is about 
one third of 
the discrepancy of $43(16)\times10^{-10}$ 
found by the recent Brookhaven experiment \cite{brook01}.
The problem is more serious 
in the case of QCD corrections.
For instance, in the calculation \cite{larin94} 
of the hadronic width of the $Z^0$,
the term of order $\alpha_s^3$ 
is more than 60 percent of the 
term of order $\alpha_s^2$ and contributes to one part in 1,000 
to the total width. 

We claim that introducing large field cutoffs leads to significantly 
improved perturbative series.
An important reference to understand the general mechanism and to 
interpret the results presented below is 
the well studied \cite{leguillou90,pernice98} integral
\begin{equation}
Z(\lambda)=\int_{-\infty}^{+\infty}d\phi {\rm e}^{-(1/2)\phi^2-\lambda 
\phi^4}\ .
\label{eq:int}
\end{equation}
If we expand ${\rm e}^{-\lambda \phi^4}$, the integrand for the order $p$ 
contribution is ${\rm e}^{-(1/2)\phi^2}\phi^{4p}/p!$ and has its maximum when
$\phi^2=4p$. On the other hand, the truncation of 
${\rm e}^{-\lambda \phi^4}$ at
order $p$ is accurate provided that $\lambda \phi^4 <<p$. 
Requiring that the peak of the integrand for the $p$-th order term 
is within the range of values of $\phi$ for which the $p$-th order truncation
provides an accurate approximation, yields the condition $\lambda<<(16p)^{-1}$.
One sees that the range of validity for $\lambda$ shrinks as one increases
the order. 
We can avoid this problem  by restricting the range of integration in 
Eq. (\ref{eq:int}) to $|\phi|<\phi_{max}$. We call the truncated integral 
$Z(\lambda , \phi_{max})$.
As the order 
increases, the peak of the
integrand moves across $\phi_{max}$ and the contribution is suppressed.
It is easy to show that the coefficients of the modified series 
satisfy the bound $|a_p|<\sqrt{2\pi}\phi_{max}^{4p}/p!$ and the modified series
defines an entire function. However, we are now constructing a perturbative
series for a problem which is slightly different than the original one.
This procedure is justified from the fact that the error is controlled 
by the inequality
\begin{equation}
|Z(\lambda)-Z(\lambda , \phi_{max})|<2{\rm e}^{-\lambda \phi_{max}^4}
\int_{\phi_{max}}^{\infty}d\phi{\rm e}^{-(1/2)\phi^2}\ . 
\label{eq:bound}
\end{equation}

We have applied large field cutoffs to the anharmonic 
oscillator with an Hamiltonian 
$H=p^2/2+\phi^2/2+\lambda \phi^4 \ .$ We use 
units such that the mass, the frequency and $\hbar$ are unity.
The method of Ref. \cite{bacus} was used to obtain 
a solution of the time-independent Schroedinger equation for 
an arbitrary value of the energy $E$. The eigenvalues are determined 
by using Sturm-Liouville theorem to monitor the ``entrance'' of the 
zeroes of the wave function in a region $0\leq \phi\leq \phi_{max}$
as $E$ increases. If $\phi_{max}$ is large enough, one obtains 
excellent 
numerical values 
of $E_n$ for $n$ not too large by requiring that the $n+1$-th zero
occurs exactly at $\phi_{max}$. 
This numerical procedure can be converted
into a perturbative expansion order by order in $\lambda$.
By taking $\phi_{max}$ large enough, namely 8,
we were able to reproduce accurately the first twenty terms
of the series for the ground state 
calculated by Bender and Wu \cite{bender69} without using diagrammatic
techniques. We have also applied large field cutoffs to 
Dyson's hierarchical model,
an approximation of lattice scalar field theory where the renormalization 
group transformation can be calculated numerically with great 
accuracy \cite{gam3rapid}. We have used 
a local Lagrangian density 
of the Landau-Ginzburg form ${-A\phi^2-\lambda \phi^4}$ such that 
when $\lambda$ = 0, the mass is unity. 
The field cutoff appears in the calculation of the Fourier 
transform of the local measure  necessary for the numerical procedure.  
The free parameter (called  $c/4$ in Ref. \cite{gam3rapid}) 
appearing in the kinetic term was chosen in a such a way that 
a free massless field scales as it would for a nearest neighbor model 
in 3 dimensions. 

\begin{figure}
\centerline{\psfig{figure=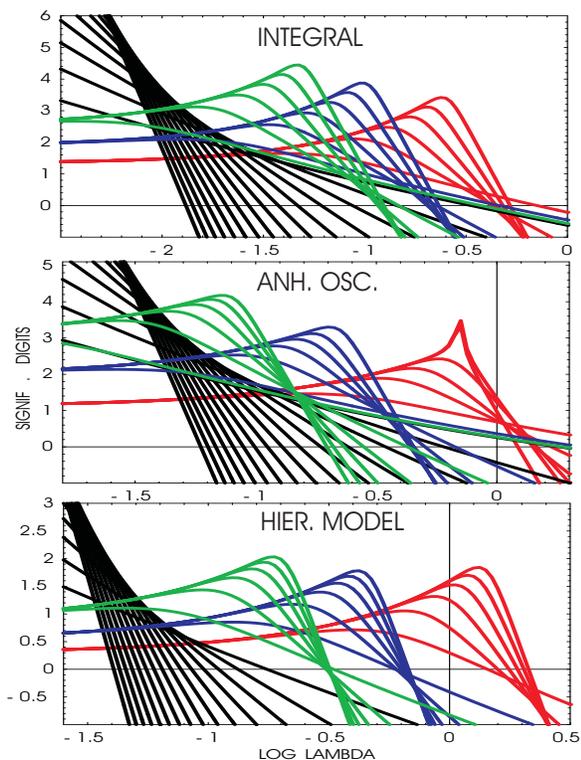,width=3in}}
\vskip10pt
\caption{Number of significant digits 
obtained with regular perturbation 
theory at order 1, 3, 5, ...., 15 (black) and
with $\phi_{max}$ = 3 (green), 2.5 (blue) and 2 (red),
at order 1, 3, ..., 11 , as a function of $\lambda$, 
for the three quantities described in the text.
Even orders have high cusps and are not displayed.}
\label{fig:two}
\end{figure}

We now present numerical results concerning the perturbative
series for $Z(\lambda , \phi_{max})$,
the 
ground state of the anharmonic oscillator and the zero-momentum 
two point function of the Landau-Ginzburg model. In Fig. \ref{fig:two},
we compare the accuracy of 
regular perturbation theory with what is obtained with
various field cutoffs. 
The curves for the modified series reach an asymptotic value on the left
and drop on the right with the same slope as regular perturbation theory but 
with an intercept on the coupling axis 
shifted to the right. In between the two regimes, 
the curve reaches a maximum. 
As the order increases, the maximum moves up and right making the 
convergence apparent in this region. In all cases,
the modified models provide an accuracy that goes far beyond the envelope
of regular perturbation theory. 
We have then compared our results with those obtained with 
Pad\'e approximants. For the clarity of the figure, we have limited ourselves
to order 3 and 4 but similar situations are observed for higher orders.
At each order, we have selected the best approximant for the series
or its Borel sum. These sums are obtained by dividing 
the $l$-th coefficient
by $\Gamma \lbrack l+1+b \rbrack$. We have followed the procedure of
Ref. \cite{baker76b}, except that at the end the inverse integral transform
was performed numerically at fixed value of $\lambda$.
We call this procedure the Pad\'e-Borel method.
\begin{figure}
\centerline{\psfig{figure=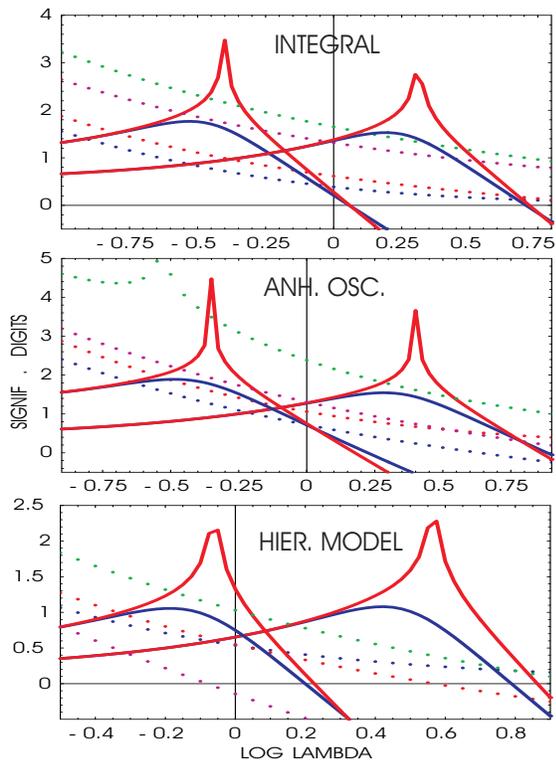,width=2.85in}}
\vskip10pt
\caption{Number of significant digits 
obtained with field cutoffs given in the text, at 
order 3 (blue line) and 4 (red line) compared to
the best approximants for the regular 
series at order 3 (blue dots) and 4 (red dots) or the best 
results obtained with Pad\'e-Borel method at order 3 (purple) and
4 (green).
}
\label{fig:three}
\end{figure}
The values of $b$ were adjusted to get 
the best possible result.
At order 4, the best approximant is $[2/2]$ in the 6 cases considered, 
but at order 
3, the situation is more complicated.
In summary, we have 
used our knowledge of the exact result to get the best possible result
for the methods to which we compare our results. Random choices of 
approximants or of the value of $b$ lead to significantly worse 
results.
The results are shown
in Fig. \ref{fig:three}. We have used field cutoffs of 1.5 and 1 for the 
integral and 2 and 1.5 for the two other models. As the field cutoff 
decreases, the curve moves right  as in Fig. \ref{fig:two}.
By comparing the three methods at the same order, we see that beyond a 
certain value of $\lambda$,
the method used here outperforms the 
two methods based on Pad\'e approximants 
within a certain range (which broadens when the coupling
increases). 
\begin{figure}
\centerline{\psfig{figure=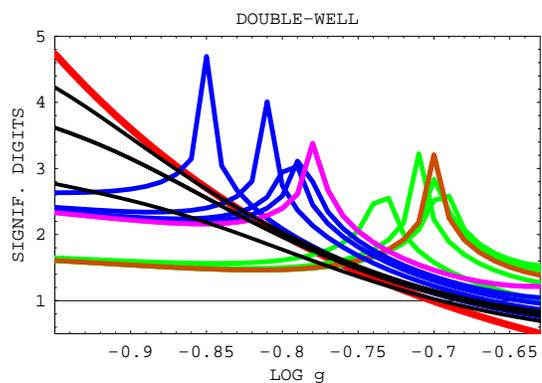,width=2.8in}}
\caption{Number of significant digits for the double-well
at order 3 to 6 for regular perturbation (black) compared to series 
obtained with $y_{min}=-3$ and $y_{max}=3$ (blue) or $y_{max}=2.5$ (green).
As the order increases, the black curves reach the one-instanton 
contribution (red) over wider regions 
to the left while the two other sets reach the accuracy level 
obtained numerically for 
$y_{max}=3$ (purple) or $y_{max}=2.5$ (brown).  }
\label{fig:four}
\end{figure} 
In all the examples considered above, the Borel sum has no singularity on
the positive real axis. One can introduce such 
singularities by adding a cubic interaction with an appropriate coupling.
However, for all examples worked out, this modification can 
be handled properly with the proposed method. 
This is due to the fact that as in the previous examples, the exponentials 
converge uniformly in a compact neighborhood of the origin, and it is 
legitimate to interchange the sum and the integral.
We report here the 
case of the double-well potential in quantum mechanics as discussed 
in Ref. \cite{brezin77}. In shifted coordinates, the potential reads
$(1/2)y^2-gy^3+(g^2/2)y^4$. By imposing the vanishing of the wave function at
$y=\pm 10$, we were able to reproduce all the significant 
digits of the first 10 coefficients 
for the ground state given in Table I of Ref. \cite{brezin77}. This series
is not Borel-summable. We have then constructed a modified series by imposing 
the wavefunction to vanish at $y_{min}< 0$ and its derivative
to vanish at $y_{max}>0$. If this prescription is 
used for numerical purposes, one obtains arbitrarily accurate results when
$g=(1/2y_{max})$ and $y_{min}$ negative enough. The numerical results are 
shown in Fig. \ref{fig:four} for values of $g$ where 
the one-instanton contribution accounts for most of the discrepancy 
obtained with the regular series. The modified series converge rapidly to 
the numerical value obtained with the corresponding $y_{min}$ and $y_{max}$.
It takes into account instanton effects and significantly improves the 
regular perturbative answer. Significant improvements have also been  
obtained for larger $g$ by decreasing $y_{max}$.

The striking resemblance among the three models appearing in Figs.
\ref{fig:two} and \ref{fig:three} suggests that, in general,
the corrections due to the field cutoffs can be 
expressed as simple one-dimensional integrals. 
The perturbative expansion of the partition function of an arbitrary lattice
scalar field theory with a large field cutoff 
can be 
obtained by writing the truncated integral at each site as 
the integral over the whole real axis minus the integral over 
$|\phi|>\phi_{max}$. Regrouping the contributions with $0, 1, \dots$ 
large field contributions, we obtain the partition function 
\begin{eqnarray}
\nonumber
Z[J]=C{\rm e}^{-\lambda \sum_x({\partial\over{\partial J(x)}})^4}
{\rm e}^{{1\over2}\sum_{ y,z}J(y)G(y,z)J(z)} \times \\ 
\label{eq:wick}
(1-A_1\sum_y 
\int_{|\phi_y|>\phi_{max}}{\rm e}^{-A_2(\phi_y-\sum_zG(y,z)J_z)^2}+\dots) ,
\end{eqnarray}
with $A_1=(2\pi G(0,0))^{-1/2}$, $A_2=(2G(0,0))^{-1}$, 
$G(x,y)$ being the two-point function at $\lambda=0$ (with no field cutoff)
and all quantities
being written in lattice spacing units. The dots in Eq. (\ref{eq:wick})
are calculable and exponentiate in the limit of 
a coarse lattice 
where the correlations among the sites are small.
In general, Eq. (\ref{eq:wick}) can be interpreted in terms of 
Feynman diagrams. 
A continuum version of this expression can be obtained by 
using a dilute-gas approximation for configurations with 
one ``lump'' of large values.
We carried the detail of this calculation in the case of the 
anharmonic oscillator using the classical configuration $\phi_{max} {\rm e}
^{-|\tau-\tau_0|}$ and adapting the arguments
of Ref. \cite{coleman}. The result for the zero-th order correction to 
the ground state reads: 
\begin{equation}
\delta E_0^{(0)} \simeq
4 \pi^{-1/2}\phi_{max}^{2}\int_{\phi_{max}}^{+\infty}d\phi{\rm e}^{-\phi^2}\ .
\label{eq:dele}
\end{equation}
This prediction fits the numerical data 
over a wide range 
of $\phi_{max}$.
We can estimate 
the optimal value of $\phi_{max}$ 
without knowing the numerical answer. 
The left part of the curves shown in Fig. 
\ref{fig:two} can be estimated by semi classical methods while the 
right part is given by the next order contribution.
In the case of the anharmonic oscillator, using the classical configuration 
mentioned above, we obtain
the error at order $n$ :
\begin{equation}
|\delta E_o (\lambda)|
\simeq\delta E_o^{(o)}{\rm e}^{-(1/2)\lambda\phi_{max}^4}+
|a_{n+1}|\lambda^{n+1}
\label{eq:err}
\end{equation}
This approximate formula fits the 
data very well if $\phi_{max}$ is not too small 
and allows a good estimate of the 
value of the coupling where the accuracy peaks.

The semi-classical calculation
performed for the anharmonic oscillator can be extended to 
other scalar theories with exponentially decaying
two-point functions and we expect an exponential control of the 
error for these models.
In the case of lattice gauge theory,
the integration over
the fields is already reduced to a compact space.
In the literature on lattice perturbation theory (with the exception 
of van Baal \cite{vanbaal91}), one usually replaces $\int dg$ by 
$\int_{-\infty}^{+\infty}dA_{\mu}^i$ since
in the continuum limit the range becomes infinite.
We claim that this lattice artifact can be used 
to obtain a  smooth truncation of the perturbative series 
as in the scalar case. Different approximations need to 
be developed to solve the massless quadratic theory with a field cutoff.
We expect that, as in the case of the double-well, this approach will lead to 
a quantitative understanding of the instanton effect.

This research was supported in part by the Department of Energy
under Contract No. FG02-91ER40664.

\end{multicols}
\end{document}